# Efficient Virtual Network Function Placement Strategies for Cloud Radio Access Networks

Deval Bhamare[¥], Aiman Erbad[¥], Raj Jain[φ], Maede Zolanvari[φ], Mohammed Samaka[¥]
[¥]Department of Computer Science and Engineering, Qatar University, Doha, Qatar
[φ]Department of Computer Science and Engineering, Washington University in Saint Louis, USA
devalb@qu.edu.qa, aerbad@qu.edu.qa, jain@wustl.edu, maede.zolanvari@wustl.edu, samaka.m@qu.edu.qa

*Abstract*—The new generation of 5G mobile services place stringent requirements for cellular network operators in terms of latency and costs. The latest trend in radio access networks (RANs) is to pool the baseband units (BBUs) of multiple radio base stations and to install them in a centralized infrastructure, such as a cloud, for statistical multiplexing gains. The technology is known as Cloud Radio Access Network (CRAN). Since cloud computing is gaining significant traction and virtualized data centers are becoming popular as a cost-effective infrastructure in the telecommunication industry, CRAN is being heralded as a candidate technology to meet the expectations of radio access networks for 5G. In CRANs, low energy base stations (BSs) are deployed over a small geographical location and are connected to a cloud via finite capacity backhaul links. Baseband processing unit (BBU) functions are implemented on the virtual machines (VMs) in the cloud over commodity hardware. Such functions, built in software, are termed as virtual functions (VFs). The optimized placement of VFs is necessary to reduce the total delays and minimize the overall costs to operate CRANs. Our study considers the problem of optimal VF placement over distributed virtual resources spread across multiple clouds, creating a centralized BBU cloud. We propose a combinatorial optimization model and the use of two heuristic approaches, which are, branch-and-bound (BnB) and simulated annealing (SA) for the proposed optimal placement. In addition, we propose enhancements to the standard BnB heuristic and compare the results with standard BnB and SA approaches. The proposed enhancements improve the quality of the solution in terms of latency and cost as well as reduce the execution complexity significantly. We also determine the optimal number of clouds, which need to be deployed so that the total links delays, as well as the service migration delays, are minimized, while the total cloud deployment cost is within the acceptable limits.

## I. INTRODUCTION

Recently, because of the explosion in the number of mobile devices, demand for new online services and consequently the data traffic has grown rapidly. With the proliferation of mobile technology, there has been a burst in the traffic originating from IoT devices, video on demand (VoD), online gaming, healthcare, and many other applications. Millions of new sensing devices and online services exchanging data have significantly contributed to this trend. It is expected that the volume of mobile data will be 1000X higher than today, and the number of connected devices will be between 10X to 100X by 2020 [3]. According to Wireless World Research Forum (WWRF), the number of connected wireless devices is expected to be 100 billion by 2025 [2]. The unprecedented growth in online services, mobile devices, as well as the data has exerted tremendous pressure on the cellular network operators to provide the connectivity to their end-users while maintaining the quality of service (QoS).

To accommodate this growth, network operators have to deploy more and more base stations to offload traffic from congested cells. Increasing the number of base stations to meet the growing user demand increases the capital expenditure (CAPEX) and operational expenditure (OPEX) for cellular network operators. More specifically, CAPEX increases as base stations are the most expensive components of a wireless network infrastructure, while OPEX increases as cell sites demand a considerable amount of power and resources to operate. However, revenues for the operators are still flat [1]. The aforementioned problem will be aggravated by the introduction of 5G networks. 5G networks will incorporate different types of heterogeneous traffic and 5G operators will be confronted with the major challenge to support a number of diverse vertical industry applications in order to expand the wireless market. Table 1 provides a summary of typical examples of such services, which illustrate the wide diversity of their associated requirements.

TABLE 1. NETWORK SERVICES AND DEMANDS

| Case | Application | Requirements |
| --- | --- | --- |
| Broadband access in dense areas | Events, games, etc. | High traffic volume, ms latency |
| Mobile users | Trains, vehicles, drones | Connectivity at high speed |
| Massive IoT | Sensors, smart devices, wearables | Low power, around 1 million connections per $km^2$ |
| Time sensitive | Health, smart grid, etc. | Redundancy, ms latency |

A novel mobile network architecture that minimizes the operational cost for network operators while accommodating such increasing heterogeneous user demands and satisfying the QoS has become a necessity. Cellular operators have started to experiment with novel networking paradigms, new ways to leverage existing equipment in new deployments, and more flexible resource planning and network managing tools. Cloud radio access network (CRAN) is a novel mobile network architecture, which has the potential to meet the above-mentioned challenges. It is based on a concept proposed by Lin et al. [8], which allows cellular network operators to share the network as well as the computational resources to balance the workload over a low-cost platform. In the CRAN, baseband processing is centralized as a virtualized baseband processing unit (BBU) pool, shared among several sites as well as operators. The idea is to virtualize BBU pools using Infrastructure as a Service (IaaS) model offered by cloud service providers. Various collaborating operators may deploy the commodity hardware at different sites, forming various cloud sites, on which the BBU functions may be collocated [5]. Resource sharing along with virtualization promotes flexible control, low cost, efficient resource usage, and support for diversified applications. In addition, communication among co-located BBUs has a lower latency, guaranteeing the QoS. Sharing of resources also results in an increased throughput by means of the statistical multiplexing gains [7, 49]. Suggested CRAN architecture is shown in Fig. 1.

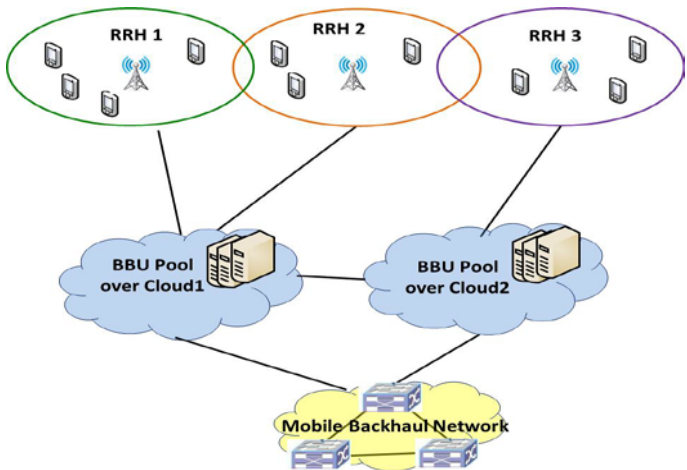

Fig. 1: Novel CRAN architecture for mobile networks.

As depicted in Fig. 1, all baseband signal processing (BBU) functions (including physical, MAC and network layers), which require most of the processing resources have been relocated from the cell site to distant locations, i.e., the clouds. On a cell site, RRH is still responsible for transreceiving radio signal, amplification of signal power and other functions [5]. In real-time scenarios, cellular network operators need to start multiple instances of the BBU services depending on the client demands or the workload. Operators confront the problem of optimally placing the service instances, considering the user demand density across multiple regions. The delays to the end-users depend on the RRH to cloud allocation as well as the service placement over the resource pool. A non-optimal allocation may result in unacceptable delays, violating the QoS and costs, hampering the advantages of the novel CRAN architecture. Hence, efficient algorithms are required to map the BBU service requirements to the available virtual resources and to minimize the end-to-end delays to the end users. The proposed algorithms are also expected to reduce the total cost of deployment to cellular network operators.

Researchers have considered the optimization problem in the context of CRANs already; with the focus on optimizing the resources, total energy or power [21, 48]. Latency, however, is equally important and faces stringent constraints in radio networks. A significant amount of work has been done in the literature for service placement considering the parameters such as latency, cost, QoS and others, especially from the context of the core telecom networks [11, 15, 27, 32-40]. The requirements of the CRAN networks, on the contrary, are significantly different compared with those of the core telecom networks [5, 21]. For example, diversified traffic on 5G networks (as explained in table 1). Also, BBU functions involve huge traffic volumes and have stringent latency requirements, in the order of a few hundreds of microseconds [5, 21, 24]. This demands high network link capacities. For example, high-definition (HD) video provided by the application service providers (ASP), such as YouTube requires a minimum of 3−4 Mbps of link capacity to satisfy the quality of service (QoS) demands [4]. Considering such stringent service level agreements (SLAs), there is a need to revisit the service placement problem from the perspective of CRANs.

In this work, we propose a scheme for optimal placement of virtual functions (VFs) in multi-cloud environments for CRANs. The proposed schemes in this work, allocate service demands at base stations in the form of VFs to the virtual machines (VMs) to minimize the response time or latency to the clients, satisfying the cost constraint, the capacity constraints and the placement constraint (due to SLAs, explained later). We have formulated the optimization model to deploy workflows of the VFs and assign service requests at base-stations to meet the service demands. We model the problem as a combinatorial optimization problem. Since the optimization model cannot solve the real-time problems within acceptable time limits due to its computational complexity, we propose a set of heuristic approaches for large networks. In this work, we implement an enhanced version of the two common approaches in the literature, which are: (1) Branch-and-bound (BnB) and (2) Simulated annealing (SA). The enhancement reduces the execution complexity of the BnB heuristic so that the allocation is faster. The proposed enhancements also improve the quality of the solution significantly. We compare the results of the standard BnB and SA schemes with the enhanced approaches to demonstrate these claims. Our aim was to develop a faster solution which can meet the latency requirements of the CRANs [8], while the performance (here, in terms of cost and latency) is not far from the optimal. Also, with the heuristic implementation, we calculate the optimal number of clouds, which need to be deployed so that the total links delays, as well as the service migration delays, are minimized, while the total cloud deployment cost is within the acceptable limits.

The rest of the paper is organized as follows. In section II, we discuss the related work in the context of the CRANs and discuss our contributions in more depth. In section III, we formulate an optimization model for the VFs placement over multiple-clouds. In section IV, we discuss the proposed heuristic approaches, and in section V, we present and discuss the results. Finally, we conclude the paper. A list of acronyms used in the paper is given in Table 8 at the end.

## II. RELATED WORK AND CONTRIBUTIONS

As mentioned by Checko et al. [4] and Lin et al. [8], the evolution of the CRAN, network operators may experience the following benefits:

(1) *Reduced cost*: Since computing resources are aggregated at a centralized location, deployment as well as maintenance cost for separate base-stations can be saved.

(2) *Increased energy efficiency*: Power consumption and load congestion can be reduced by dynamically allocating the resources and allocating the services over the shared pool, energy efficiency may be improved significantly.

(3) *Improved spectrum utilization*: CRANs enable sharing of channel state information on each base station-mobile station link, traffic data, and control information of mobile services among cooperating base stations, resulting in an improved spectrum utilization [5, 9].

(4) *Improved resource utilization*: Since computers and other resources are shared, overall resource utilization can also be significantly improved.

(5) *Scalability*: An RRH site can be easily deployed or undeployed as per the need, without worrying about the installation of the whole base-station. Such new sites can be multiplexed with the existing centralized BBUs [5, 47].

Considering the benefits of the CRAN architecture, researchers have already started to investigate these challenges associated with this novel platform. For example, Hadzialic et al. present an overview of all known techniques to realize a CRAN network [14]. Wu et al. [7] present a novel logical structure of CRAN that consists of a physical plane, a control plane as well as a service plane and emphasizes the advantages of the CRAN architecture. The authors propose a coordinated user scheduling algorithm and a parallel optimum pre-coding scheme using cloud computing platforms. Dario et al. [16] introduce the concept of RANaaS (Radio Access Network-as-a-Service) as a flexible architecture based on the centralized processing. Checko et al. [5] provide a technology review for this novel platform, focusing on its advantages. Going a step further, Qian et al. [1] propose a super-base station based centralized approach for 5G networks. The authors also acknowledge the advantages of the CRAN architecture in their work. Navid [8] investigates three critical issues for the cloudification of the current radio access networks. The author analyzes resource, latency, and capacity requirements for the baseband processing units. Liu et al. in [13] demonstrate that energy efficiency of large-scale small cell networks is higher compared to massive multiple-input multiple-output (MIMO) systems. Mikhail et al. [50] discuss cooperative radio resource management approaches in heterogeneous CRANs. Park et al. [51] have focused on massive MIMO perspective and related operations for the partially centralized CRANs.

Peng et al. [21] discuss recent advancements in the field of CRANs. They also provide a survey of technological features and the core principles of the heterogeneous CRANs in [20, 24, 45]. Rost et al. [25] provide an overview of the cloud technologies for 5G radio access networks and discuss the advantages of CRANs. Pang and Zhang discuss the recent advances in the field of fog radio access networks [47]. A significant amount of work in the literature concerns the virtualization of core telecom as well as network functions using software-defined networking (SDN) and virtualization. Our previous work [3] has considered the placement problem from the perspective of service function chaining (SFC) of application layer as well as network layer services. Virtual function or virtual machine placement problem has been considered widely in the literature.

Various optimization models for resource allocation in radio networks, as well as core networks, have been proposed along with the heuristic approaches. For example, optimal energy-efficient power allocation schemes for radio networks by Weng et al. [28]. Sigwele et al. [48] have proposed energy-efficient CRANs by cloud-based workload consolidation for 5G networks. A virtual machine placement problem in micro-cells while implementing SLA constraints has been considered in [12]. In [50] authors propose the novel caching scheme to solve the problem of congestion in backhaul links. Park et al. [51] propose a joint optimization scheme to tackle the problem of maximizing the delivery rate in fog RANs. Virtual machine placement and service placement have been studied in the literature as well. For example, works presented in [29-40]. A multi-cloud virtual function distributed strategies for dynamic NFV platform has been proposed in our previous work [15].

Besides academia, cellular network operators have extensively studied CRAN architecture as well, such as Alcatel-Lucent [17], Huawei [18], Nokia Siemens Networks [19], China Mobile Research Institute [6] and many others. Given the importance of CRANs and their stringent latency requirements, we argue that it is mandatory to implement sophisticated algorithms for automated service delivery in the context of multi-cloud based RANs to fully leverage the distributed computing opportunities on the Internet. In this work, we aim to achieve the following objectives to address the service placement problems in the context of CRANs:

(1) We solve the problem of automated and optimal service placement for the multi-cloud domain in the context of CRANs.
(2) We propose the optimization model to minimize the overall latency while placing the BBU services over the centralized cloud and satisfying the cost as well as resource constrains.
(3) We also implement enhanced versions of the two common heuristics in the literature, which are: (i) Branch-and-bound (BnB) and (ii) Simulated annealing (SA). The enhancement reduces the execution complexity of the BnB heuristic with the improved solution quality.
(4) Finally with the heuristic implementation, we calculate the optimal number of clouds, which need to be deployed so that the total links delays, as well as the service migration delays, are minimized, while total cloud deployment cost is within the acceptable limits. We also validate our results using the results from NS3 for a similar setup.

In the next section, we propose our optimization model for the deployment of BBU virtual functions over the available clouds and a set of virtual machines, so that the total end-to-end delays for all the services are minimized while satisfying the cost and capacity constraints.

### III. OPTIMAL DISTRIBUTION OF BBU SERVICES

In this section, we set up the problem of minimizing the overall response time in a multi-cloud RAN (CRAN) scenario as a combinatorial optimization problem. For typical batch-processing cloud computing applications, delay ranging in the tens of milliseconds is acceptable, for CRANs the expected delay should be less than 0.5 milliseconds [5]. This stringent delay requirement mandates the transport networks to not only support high bandwidth and be cost efficient but also to support strict latency and jitter requirements. Hence, the goal of the optimization model presented is to minimize the response time or latency to the services, while satisfying other constraints such as the cost, placement, and capacity constraints.

We have formulated the optimization model to deploy the workflows using the virtual network functions (VNFs) and assign client requests to these workflows to meet the service demands. A service request is nothing but a resource request vector by the individual VNF that comprise the workflow. We introduce a 3D model for service requirements and VM capacities that are necessary for multi-cloud RAN scenario, that is, CPU, storage, and network capacity. For more detailed discussion on the workflows, their characteristics and service requests, readers are requested to refer to our previous works such as [14, 52].

In our model, link delays among base stations and the cloud are considered because such delays are of significant importance in cellular networks. We formulate the problem as an integer linear programming (ILP) optimization. The list of variables used in the ILP is given in Table 2. Let $G = \{V, E\}$ be a graph to represent the network in consideration, where $V$ is a set of nodes representing the base stations and cloud nodes in the network and E is set of the edges such that $E \subseteq V \times V$. To reduce the computational complexity of the optimization model, we compute the path between every pair of the nodes in the topology in advance, mapping paths to the links. The mappings are stored in the set $\xi$. The total number of sites that can be selected for deployment of the clouds is given as an input to the optimization model and depends on the cost threshold, that is, $\Gamma$.

TABLE 2
PARAMETERS FOR INTEGER LINEAR PROGRAM (ILP)

| Type | Symbol | Definition |
|---|---|---|
| Indices | $i, j, k$ | Iterators for nodes in the topology such that $i, j, k \in |V|$ |
| | $l$ | Iterator for virtual machines that $l \in L$ |
| | $m$ | Iterator for service instances such that $m \in M$ |
| | $p$ | Iterator for the paths such that $p \in P$ |
| Input Constants | $V$ | Set of nodes in the topology |
| | $\psi_j$ | The arrival rate of packets at $j^{th}$ node (Poisson distribution). |
| | $\Psi_j$ | The processing rate at $j^{th}$ node (Poisson distribution) |
| | $C_j$ | Computational delay at the $j^{th}$ cloud. Clouds are modeled using an M/M/1 model. |
| | $\lambda_{ij}$ | The arrival rate of packets on the path $(i, j)$. |
| | $\mu_{ij}$ | Processing rate of packets on the path $(i,j)$ (deterministic) |
| | $T_{ij}$ | Total delay on $(i, j)^{th}$ link/path to transmit one byte. Link queues are modeled using an M/D/1 model. |
| | $B_{ij}$ | The bandwidth of the link between the $i^{th}$ and $j^{th}$ node. Value is 0 if no direct link between $i$ and $j$ |
| | $K_j$ | The capacity vector of the $j^{th}$ node (3D vector). |
| | $\kappa_l$ | The capacity vector of $l^{th}$ VM (3D vector). |
| | $\delta_l$ | Demand vector for $l^{th}$ VM (3D vector). |
| | $\Delta_i$ | Demand vector for $i^{th}$ BS per byte of traffic (3D vector). Value is 0 if the $i^{th}$ node is a cloud node. |
| | $W_i$ | Traffic generated by $i^{th}$ BS in a number of packets. Value is 0 if the $i^{th}$ node is a cloud node. Each packet size is assumed to be 500B. |
| | $S_i^l$ | A 2-dimensional $M \times |V|$ matrix. Value is 1 if $l^{th}$ function can be placed at $i^{th}$ cloud location based on the SLAs, otherwise 0. |
| | $\Psi_m$ | The delay constraint for service $m$, such that SLAs are met. |
| Variables | $I_{jm}^l$ | Instance matrix indicating number of instances of $l^{th}$ VM which are installed on $j^{th}$ node for $m^{th}$ service request. |
| | $A_{ij}^m$ | Allocation matrix. Value is one if $i^{th}$ node (BS node) is assigned to $j^{th}$ node (cloud node) for $m^{th}$ service request otherwise 0 |

We assume that the set of base stations (BSs) and clouds are disjoint sets. Each cloud site $i$ has zero value for $W_i$, that is, no request flows are getting generated in clouds and only base stations can generate such flows. Similarly, each BS site $i$ has zero value of $K_i$, that is, user sites do not have any processing capacities. A vector matrix $K$ represents the capacities of the cloud sites, with each element of the matrix being a triplet $K_i = [K^1_i, K^2_i, K^3_i]$ is the capacity of cloud $i$. As mentioned earlier, we are referring to a 3D vector to represent the capacity, that is, CPU, Storage and, Network Capacity. Let $W$ be the matrix to represent the volume of traffic originating from the BS sites, that is, $W_i$ is the traffic getting generated at BS node $i$. $K_i = 0$ and $W_i > 0$ indicates that the site $i$ is a base station site. However, $K_i = 0$ and $W_i = 0$ indicate that the node is just a routing node. We assume that service requirements are directly mapped to virtual machines (VMs) for their installations. For simplicity, the mapping is assumed one-to-one, hence, we may be using both the terms interchangeably. $\kappa_l$ is the vector representing capacity required for the $l^{th}$ VM. Let $\delta_l$ be the demand vector of $l^{th}$ VM and $\Delta_i$ be the demand vector for the $i^{th}$ client. For the cloud node $\Delta_i = 0$.

It may be noted that more than one instance of a VM may be deployed at any deployment site depending on the processing capacity of the VM and total traffic demand getting generated at the site. Let $I_j^l$ be the instance matrix representing how many instances of a VM $l$ need to be deployed at site $j$. Let $A$ be an allocation matrix such that $A_{ij}^l = 1$ if a BS node $i$ is assigned to the cloud at node $j$. Note that $A_{ii}^l = 1$ means node $i$ has been assigned a client request. In other words, $l$ instance has been deployed on a cloud at node $i$. If $N$ is the number of the total nodes and $L$ is the total number of VFs to be deployed, then $A$ can be given as:

$$A = \begin{bmatrix} A_{11}^1 & \ldots & A_{1N}^1 \\ \ldots & \ldots & \ldots \\ A_{N1}^1 & \ldots & A_{NN}^1 \end{bmatrix} \ldots \begin{bmatrix} A_{11}^L & \ldots & A_{1N}^L \\ \ldots & \ldots & \ldots \\ A_{N1}^L & \ldots & A_{NN}^L \end{bmatrix}, \forall A_{ij}^l \in (0,1) \quad (1)$$

As mentioned earlier, the computing systems are modeled as M/M/1 queues. Similarly, the links are modeled as M/D/1 queues with large buffers, up to 1 GB. The delays in the links are as given in Equation (2). We note that $\lambda_{ij}$, which is the link load, is a function of total flows passing through the link $(i, j)$. Also, $\psi_j$ is the arrival rate of packets and $¥_j$ is the process rate of packets at the $j^{th}$ cloud.

$$C_l = 1/(1- \psi_l/¥_l)$$

$$T_{ij} = \frac{1}{2\mu_{ij}} \times \frac{2-(\lambda_{ij}/\mu_{ij})}{1-(\lambda_{ij}/\mu_{ij})} \quad (2)$$

***Constraints***: We now discuss the constraints of the optimization model:

1. Cloud capacity: The maximum number of instances of a VM, which may be deployed in a given cloud, is bounded by the capacity of that particular cloud and demands of the VM. In other words, the summation of the demands of all VMs installed in a cloud $j$ should be less than or equal to the capacity of the cloud $j$.

$$\sum_{m=1}^{M} \sum_{l=1}^{L} I_{jm}^l \times \delta_l \leq K_j \,\forall j \in |V|, m \in M \quad (3)$$

2. VM Capacity: The minimum number of VMs that need to be deployed on a particular cloud for a particular service is bounded by a fraction of the total client traffic from all the sites assigned to that particular cloud. It means that the sum of the demands of BSs assigned to a particular service $m$ at a particular site $j$ should be less than or equal to the total capacity of all instances of that particular VM $l$ at site $j$.

$$\sum_{i=1}^{|V|} A_{ij}^m \times \Delta_i \times W_i \leq I_{jm}^l \times \kappa_l, \forall j \in |V|, m \in M \quad (4)$$

3. Link Delays: This constraint models the total path loads as a function of total traffic passing through the path between node pair $(i, j)$. $\lambda_{ij}$ is the total load across the path between $(i, j)$, and is a function of total flows passing through the path between $(i, j)$. Please note that, to reduce the complexity of the model further, paths are pre-calculated. Generally, the backhaul links between the BS and the first cloud (or the first routing element) are very high capacity links [46]. This is because the cellular network providers lay such links and the first cloud is generally close to the base station. However, the links between the clouds are generally laid by ISPs and have comparatively lower capacities. Such inter-cloud links may become bottlenecks. Hence, we consider only such links and ignore the links between the BSs and the first cloud. $\lambda_{ij}$ may be given as shown in Equation 5.

We modify the term $\lambda_{ij}$ for this equation to $\lambda_{ij}^p$ to indicate the delays on the end-to-end path between node pair $(i, j)$. For the sake of clarity, we replace the term $i$ with the term $i'$ to indicate that the first node on the specified path is not the $i^{th}$ BS, however, it is the first routing element or the first cloud. Also, term $A_{ij}^m$ in the equation makes sure that $i^{th}$ BS is allocated to cloud $j$ for service request $m$. Term $A_{jj}^m$ validates the integrity constraint, that is, node $j$ is a cloud. $\xi_{ij}^p = 1$ indicates that the path $p$ has been chosen for the communication between the end nodes $(i, j)$,

$$\lambda_{i'j}^p = \sum_{i \,\varepsilon\, |V|} \sum_{m \,\varepsilon\, M} \xi_{i'j}^p \times W_i \times (A_{ij}^m \times A_{jj}^m), \ \forall \, i'j \in |V| \quad (5)$$

4. Queuing Constraints: For the queuing systems to be stable, following two constraints need to be satisfied. That is, processing rate should be greater than or equal to the arrival rate, at both, links as well as clouds.

$$\lambda_{ij} \leq \mu_{ij} \text{ and } \psi_j \leq ¥_j \qquad \forall i, j \in /V/ \quad (6)$$

5. Cost Threshold: The total cost of deployment, $\Gamma$, is an input for our ILP. $\Gamma$ varies from $\Gamma_{min}$ to $\Gamma_{max}$. $\Gamma_{min}$ generally is 1. However, we allow the possibility of starting with another feasible number. The cost associated with a single cloud is proportional to the total resources installed at that site. We iterate through the matrix $A$ to count cloud sites and find the cost to make sure that the total cost is less than or equal to $\Gamma$.

$$\sum_{m=1}^{M}\sum_{i=1}^{|V|} A_{ii}^m \times I_{im}^l \leq \Gamma \qquad (7)$$

6. SLAs for Response Time: Depending on the service type, the scheduler at BBUs may want to limit per-packet delays for its various service request. This also avoids starvation of a particular traffic flow due to limited resources. This constraint depends on the final optimization function for total delays. Let $\Theta_{im}$ be the total delay for the $m^{th}$ service at the $i^{th}$ BS. The constraint can be modeled as follows [15].

$$\Theta_{im} \leq \Psi_m, \forall\, m \in M,\, i \in |V| \qquad (8)$$

7. Integrity Constraint: As mentioned earlier, we assume that the set of users and clouds are disjoint sets. Hence, we need to make sure that the user requests are forwarded to cloud nodes only (and not to the other client nodes). It is ensured with the help of the following constraint:

$$A_{ij}^l \leq A_{ii}^l, \forall i, \quad j \in |V|, l \in L \qquad (9)$$

***Optimization Function***: We seek to minimize the total response time to the base stations in the network. Delays are divided into two categories: transmission delays associated with links and computational delays associated with the clouds. Then we multiply the term with the transmission delay between nodes $i$ and $j$ as well as the computational delay at node $j$ ($T_{ij}$ and $C_j$, respectively). We formulate the optimization function as follows.

Minimize:
$$\sum_{m \in M}\sum_{i \in |V|}\sum_{j \in |V|} A_{ij}^m \times A_{jj}^m (T_{ij} + C_j) \qquad (10)$$

***Linearization of ILP***: We formulate an optimization function as shown in (10) above. However, we notice a non-linearity in the equation due to the multiplication of $A_{ij}^m$ and $A_{jj}^m$. To remove the non-linearity, we introduce another variable $\Phi_{ij}^l$ such that [10, 15]:

$$\Phi_{ij}^l = 1 \text{ iff } A_{ij}^m \text{ and } A_{jj}^m = 1, \text{ otherwise } 0 \qquad (11)$$

satisfying the constraints below:
$$\Phi_{ij}^l \leq A_{ij}^m \text{ and } \Phi_{ij}^l \leq A_{jj}^m \qquad (12)$$
$$\Phi_{ij}^l \geq A_{ij}^m + A_{jj}^m - 1 \qquad (13)$$

The optimization function may be re-written as:
$$\sum_{m \in M}\sum_{i \in |V|}\sum_{j \in |V|} \Phi_{ij}^m (T_{ij} + C_j) \qquad (14)$$

The computational complexity of the optimization model is very high. We note that $A$ is a 3-dimensional matrix. Due to the term $A_{ij}^m \times A_{jj}^m$, the total complexity of the ILP is $O(V^4 M^2)$, where $V$ is the total number of user nodes and M is the total number of service requests. As $M << V$, the complexity may be written as $O(V^4)$, which is still very high.

Due to this high computational complexity, application of this optimization may be restricted to small data sets. Hence, we propose heuristic approaches in the next section to solve real time problem for a large number of users. In particular, we implement and compare branch-and-bound (BnB), with and without suggested enhancements as well as simulated annealing (SA) with a different number of iterations.

IV. PROPOSED HEURISTIC SOLUTION

In Section III, we proposed an optimization model for the placement of the BBU services (or VFs), which are deployed over the VMs, across multiple clouds. In CRANs, a single network operator may have a set of services (VFs) to be deployed, and every VF has specific requirements such as computational power, storage capacity, and network bandwidth capacities. Each VM may host a specific set of VFs depending on its computational, storage and network capabilities as well as other constraints such as delays.

Locating an optimal cloud and a VM from a given pool to minimize the number of VMs so that all the instances of the VFs can be satisfied is an NP-complete problem. It can be reduced to the "*Set Cover*" problem in polynomial time. The optimization problem has extremely high time complexity, and, hence, it cannot be used to solve larger real-time problems. In this section, we describe a set of heuristic approaches for this twofold problem of mapping the service requests to VFs and VF deployment on the preselected VMs across multiple clouds. We introduce a 3D model for service requirements and VM capacities that are necessary for multi-cloud RAN scenario, that is, CPU, storage, and network capacity. Vector matrix *C* represents the capacities of the VMs in a vector format with $C_i = [C^1_i, C^2_i, C^3_i]$ being the capacity of VM $i$. Each VF in a service workflow demands some minimum network infrastructure to be able to communicate across the clouds with other VF with a given set of SLAs to be met.

Branch and bound approach and a simulated annealing approach are the two important heuristics proposed in the literature to solve the combinatorial optimization problem. In this work, we consider these two approaches for our implementation. We implement an enhanced version of the branch and bound approach, as explained later in the section. The enhancement has been provided regarding the time complexity of the BnB heuristic, so that the allocation is faster, while the quality of the solution is significantly improved. We compare the results of the standard BnB and SA schemes with the enhanced approaches to demonstrate the statements above. We have aimed to develop a quicker solution to minimize the total response time to the base stations in the network of the RANs [8], while the performance (here, cost) is not far from the optimal. All the heuristics start with the static case, that is, service requests are known in advance. Then, heuristics continue with the dynamic case to handle service requests as they arrive. Below we explain our proposed approaches for allocation of the VFs to the VMs on the selected cloud.

*Branch and Bound Approach*: We provide an improvisation to the standard branch and bound approach by sorting the data structures, which store the cloud as well as VM capacities in advance. We sort the lists in ascending as well as descending order of the remaining capacities, as explained later in this sub-section. We name the modified approaches as a branch and bound-sorted descending (*BnB-SD*) and branch and bound-sorted ascending (*BnB-SA*), respectively. We implement the basic branch-and-bound approach as well and observe significant improvement in the solution quality as well as the overall execution time, as demonstrated in the next section. The branch and bound approach begins with generating the branches for the possible solution from the start position. BnB algorithm searches the complete solution space for a given problem for the best solution. However, the explicit enumeration is normally not feasible due to the exponentially increasing number of potential solutions. The unexplored subspaces are represented as nodes in a dynamically generated search tree, which initially contains only the root. Each iteration of BnB algorithm processes one such node. The iteration has three main components: (1) selection of the node to process, (2) bound calculation, and (3) branching. We limit the search space by sequentially iterating through the possible deployment options using the data sets.

We maintain two separate lists of the available cloud nodes and a list of available VMs in each cloud. We sort the datasets in advance to further reduce the execution time and improve the solution quality. For each BS, there is a separate list for *k-shortest* paths to each cloud are calculated in advance. For each path, remaining link capacities and total delays from the respective BS to the clouds are calculated and stored. The lists are sorted as well, in ascending order of the total delays. The first cloud in the list, which satisfies the latency and capacity requirements, is selected for deployment. After allocation, the remaining cloud and link capacities for that path are updated accordingly. Each cloud also has its own list of the VMs deployed on it, which is sorted based on the remaining capacities of the VMs. Again, the first VM, which satisfies the capacity requirements, is considered for deployment. We implement three versions of the BnB approach, as mentioned earlier. Algorithmic steps for the proposed approach are given in Table 3.

TABLE 3. STEPS OF BNB(SA & SD) HEURISTIC APPROACHES

1. Iterate through the service requests
2. Select the request sequentially (or select the first arrived service).
3. $L_C$: List for the available cloud node configurations
4. $L_{VM}$: List for already installed VM instances with the remaining capacities.
5. $L_C$ → Sort ascending(BnB-SA) or descending (BnB-SD)
   $L_{VM}$ → Sort ascending (BnB-SA), descending (BnB-SD)
6. $M(i \times k)$: matrix, for total $i$ clouds in the topology and $k$-shortest paths for each cloud.
7. The rows are sorted which can be treated as a list separately and is denoted as $L^i_j$ where $i$ is the cloud and $j$ is the BS index.
8. *For each (BS j)*
   *for each (node i)*
   $L^i_j$ → $Sort(M_i)$
9. $M$ is updated periodically as per the traffic conditions in the network.
   $M_{ij} = \frac{1}{2\mu_{ij}} \times \frac{2-(\lambda_{ij}/\mu_{ij})}{1-(\lambda_{ij}/\mu_{ij})}$
10. The first cloud in the sorted list of the BS $L^i_j$, which meets the latency and capacity requirements, is selected.
11. After allocation, paths and the lists are updated. Remaining bandwidth between node i and j is updated as:
    $(B_{ij})^R = B_{ij} - (\Delta^{mn}_i \times W_{ij})$
    (Note that superscript $^R$ indicate the remaining capacities and $^n$ indicates the network demand of the service request).
12. The first VM instance in the sorted list $L^c_v$ of the selected cloud, which meets the capacity requirements to accommodate the request is chosen for deployment. Remaining VM capacities are updated as:
    $(\kappa_l)^R = \kappa_l - (\Delta^m_i \times W_{ij})$
13. If no such VM is found, the new VM instance is launched, and lists are updated. The remaining cloud capacities are updated as:
    $(K_j)^R = K_j - \delta_l$
14. After the service completion, VM resource are freed and remaining VM capacities are updated as follows:
    $(\kappa_l)^R = \kappa_l + (\Delta^m_i \times W_{ij})$

*Simulated Annealing*: Simulated annealing is a probabilistic technique to approximate the global minimum solution. Optimization of a solution involves evaluating the neighbors, which is a random process. In this case, the solution is accepted if all the latency and capacity constraints are within the acceptable range, with probability one. Selecting the number of iterations is an important step and may affect the solution quality significantly. In this work, we select two extreme numbers of replications and observe the improvements in the solution quality against the execution complexity. Steps of the SA approach are given in table 4.

TABLE 4. STEPS OF SA HEURISTIC APPROACH.

1. Iterates through the service request and select the request sequentially or selects the first arrived service).
2. $L_C$: List for the available cloud node configurations
3. $L_{VM}$: List for already installed VM instances with the remaining capacities.
4. For each *(BS j)*
   *For each (node i)*
   $L^i_j$ → $Sort(M_i)$
5. $M_{ij} = \frac{1}{2\mu_{ij}} \times \frac{2-(\lambda_{ij}/\mu_{ij})}{1-(\lambda_{ij}/\mu_{ij})}$
6. A selection process is initiated, and a random solution is selected from the available candidate solutions.
7. The selection process is repeated *Y* times, and the best solution is selected, such that,

*Shorter runs* $¥ = \sqrt{number\ of\ total\ requests}/5$
*Longer runs* $¥ = \sqrt{number\ of\ total\ requests}$

8. Remaining bandwidth between node *i* and *j* is updated as follows:
$$(B_{ij})^R = B_{ij} - (\Delta^{mn}_i \times W_{ij})$$
9. The candidate solution, which meet the capacity requirements to accommodate the request, is chosen for deployment. Remaining VM capacities are updated as follows:
$$(\kappa_l)^R = \kappa_l - (\Delta^m_i \times W_{ij})$$
10. The selection process is repeated $¥$ times, and the best solution is selected.
11. If no such VM is found, the new VM instance is launched, and lists are updated. The remaining cloud capacities are updated as follows:
$$(K_j)^R = K_j - \delta_l$$
12. After the service completion, VM resource are freed and remaining VM capacities are updated as follows:
$$(\kappa_l)^R = \kappa_l + (\Delta^m_i \times W_{ij})$$

While considering the requirements for a particular VF, we allow it to run on a VM even if a VM cannot satisfy its complete CPU and Network requirements but instead can satisfy some percentage of it. This might degrade the VF performance, but it might perform the task if degradation is within allowed limit, rather than simply denying the VF request. However, the storage requirements need to be satisfied completely for a VF to be started. It may be justified since a network operator may prefer a service to be a bit slower rather than not running at all if allowed by the SLAs. We have assumed the percentage degradation level to be 20% (i.e., $\lambda = 0.2$), though this parameter is configurable and depends on the end-user demands or service level agreements. We repeat the above steps for each heuristic until all the services of a given network operator are deployed. This procedure is repeated for all network operators.

To conclude the analysis, we now analyze the time complexity of the proposed heuristics. As mentioned earlier, for modified BnB approach, we maintain two separate lists, one for the VMs and another for the clouds. The lists are sorted in (either ascending or descending) order of the total capacities and remaining capacities respectively. This can be done with standard sorting algorithms in time complexity of $O[(V+C) \times logV]$ where $V$ is the number of VMs and $C$ is the total number of clouds in the system [20]. Generally $V >> C$ and hence for brevity, we will be considering the terms containing $V$ only. To find a match for the incoming service-requests (or VFs) binary search is implemented to fasten the search since the lists are already sorted as per the capacities. Binary search can be performed in time complexity of $O(log V)$ [20]. If $M$ is the total number of the service requests, the total complexity of the proposed heuristic sums up to $O(V \times logV) + O(M \times logV)$. This can be written as $O((M+V) \times logV)$. If we assume sets $M$ and $V$ have approximately the same size, the final complexity is $O(2M \times logM)$ or $O(M \times logM)$, where M is the problem size. Without sorting, the simple BnB approach has a time complexity of $O(M \times V)$ or simply $O(M^2)$. We also observe that the time complexity of the SA approach depends on the number of iterations or the randomization factor. If $¥$ is the number of iterations, then the total delays can be given as $O(MV \times ¥)$, or simply $O(M^2 \times ¥)$. If $¥$ is small compared to the input size, the time complexity is of the order $O(M^2)$. However, if the value of $¥$ is comparable in the problem size, then the complexity of the SA approach rises to $O(M^3)$.

In the next section, we describe our experimental setup and the results obtained to evaluate the performance of the proposed heuristics.

V. RESULTS AND ANALYSIS

In this section, we first explain the experimental setup and then discuss the results obtained to demonstrate the superiority of the enhanced BnB approaches.

*V.I Experimental Setup*: For the experimental setup, we have considered a closed-loop system. Each request from a base station is assumed to be a set of 1000 data packets. For this set of packets or one request, a single reply is sent back from the cloud to the base station as an acknowledgment for the request completion. The next request is sent only after the reply to the previous one has been received. Every BS sends a predefined amount of data for every service, selected randomly from a predefined set. Depending on the desired rate of transmission, the base station sends data at a specific rate. For example, if the $k^{th}$ request has 10 GB of data to send, then that particular BS will generate $10^7$ packets of 1024 bytes each. Also, we consider link delays and computational delays in our model. As indicated earlier, the link queues are modeled as *M/D/1* (single server/Poisson arrival/deterministic service times) and server queues as *M/M/1* (single server/Poisson arrival/exponential service times) based on the statistical analysis [3, 14]. We vary the number of available clouds and observe the variations in the total delays as the number of clouds increase. Total delay on $(i, j)^{th}$ link/path to transmit one byte can be given as:

$$\Gamma_{total} = \sum_{x=1}^{L} \frac{1}{2\mu_x} \times \frac{2-(\lambda_x/\mu_x)}{1-(\lambda_x/\mu_x)}$$

Similarly, computational delays at $j^{th}$ cloud can be given as:

$$C_j = 1/(1- \psi_j/¥_j).$$

We generate sample CRAN topology as shown in Fig. 2. We vary the number of BS and the number of clouds as per the specified parameters. As shown in the figure, there are 50 BS, connected to five clouds with aggregation and core routers.

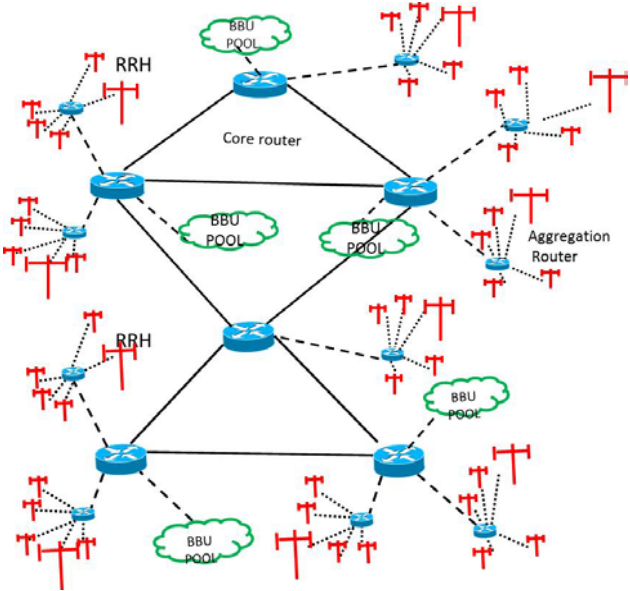

Fig. 2: Sample topology considered for heuristic implementation.

We now explain the experimental parameters we assumed for the execution of the heuristics so that the BBU requirements are closely matched. We carry out the experiments until the total number of requests reaches 10K, displayed along the X-axis on the graphs. BBU service requirements have been taken from [5] and have been mapped to the available VM configurations. For example, generally, BS functionality may be divided into the four categories such as physical layer, Lower MAC layer, Upper MAC layer, and Network Layer functionality. The functional division and their resource requirements for 10 Gbps traffic are given in table 5. Please note that these requirements may vary in the real-time scenarios.

TABLE 5. BS FUNCTIONALITY DIVISION.

| Functionality | vCPUs | NW (Gbps) |
|---|---|---|
| Physical | 2 | 5 |
| MAC-Lower | 4 | 2 |
| MAC-Upper | 6 | 1.5 |
| NW | 8 | 0.5 |

VM configurations and costs in these experiments have been taken from Amazon EC2 [44]. A sample table for the available VM configurations and their costs is given in Table 6. Inter-service arrival times have bene assumed to be exponentially distributed [3, 15].

TABLE 6. SAMPLE VK CONFIGURATIONS TAKEN FROM [4].

| Model | vCPUs | Memory (GB) | NW (Gbps) | Cost ($/HR) |
|---|---|---|---|---|
| 2×Large | 8 | 61 | 5 | 0.532 |
| 4×Large | 16 | 122 | 10 | 1.064 |
| 8×Large | 32 | 244 | 10 | 2.128 |
| 16×Large | 64 | 488 | 20 | 6.669 |
| 32×Large | 128 | 1952 | 20 | 13.338 |

***V.II Results and Discussion***: We now discuss the quality of the solution obtained with the proposed heuristics and compare their performances. As shown in Fig. 3, simulated annealing (SA) with small number of iterations takes the least time to complete, which is expected. For shorter runs, we performed $\sqrt{number\ of\ total\ requests}/5$ iterations. For longer runs we performed up to $\sqrt{number\ of\ total\ requests}$ iterations, which are in accordance with the input size. . SA with longer runs takes significantly larger time to complete the execution. The subsequent graphs discussed the improvement in the solution quality with the increase in the runs. As observed, all the three BnB approaches take approximately similar time, which lay in between the two SA extremes. For example, BnB approaches take approximately 17,000 milliseconds to complete the runs for around 8,000 requests. SA–shorter takes 5,000 ms while the SA–longer takes more than 30,000 ms for the same number of service requests.

In the next graph, we compare the heuristics based on the number of the satisfied service requests, with limited resources installed on the clouds. We classify the resources as fixed resources and flexible resources (such as VMs), which may be deployed in the clouds. However, there is an upper limit on the total resources which can be deployed (we have assumed the upper limit to be 50,000 in terms of the normalized resource numbers). As seen from Fig. 4, with the given total resources, BnB-SA performs the best with no request drop until around 8,400 requests. The maximum service drop is 2,000 at 10,000 service requests. That is, with the given resources, BnB-SA could accommodate as many as 8,000 requests, which is the highest. SA approaches performed the worst with drops starting at 5,400 requests for small runs and drops starting at 5,600 requests for long runs. However, it is still far below the best. For BnB-SD and BnB normal approach, the numbers are 7,000 and 6,500, respectively.

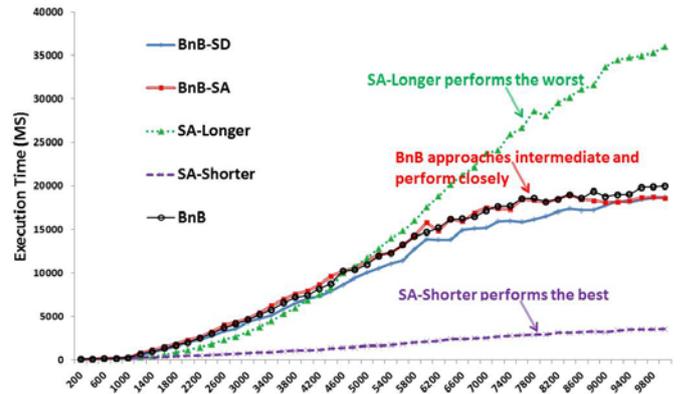

Fig. 3: Execution time comparison.

In Fig. 5, we have plotted the graphs for service migration instances, which occur when already deployed services are re-adjusted in order to accommodate the new incoming service requests. More service migration instances mean more re-adjustments and eventually more delays to the responses. Hence, we seek to minimize these migration instances. Again, BnB-SA performs the best, with only 600 maximum migration instances under the heaviest load. BnB-SD on the contrary resulted in 1200 migrations. As observed in Fig. 6, BnB-SA also has the lowest overall delays. Delays gradually increase for SA-shorter, BnB, BnB-SD and then SA-longer. Since BnB-SA has the lowest service migration instances and intermediate execution time, it has the lowest total delays as well.

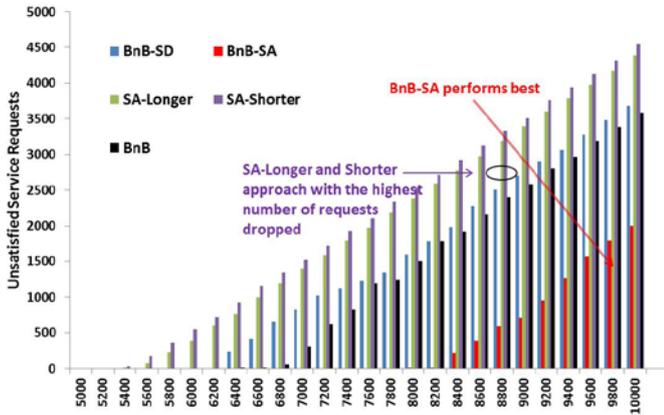

Fig. 4: Unsatisfied number of user requests.

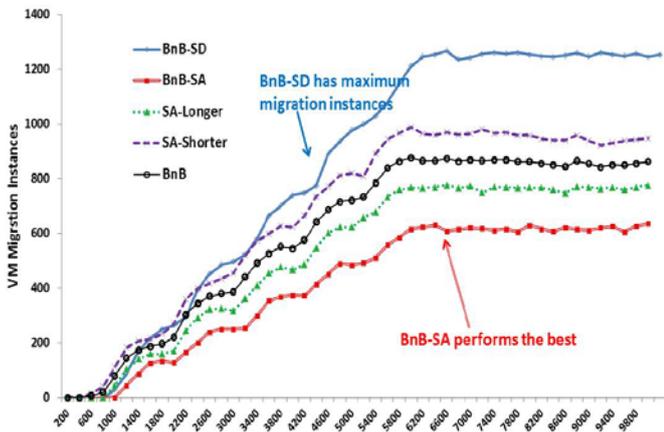

Fig. 5: Service migration instances.

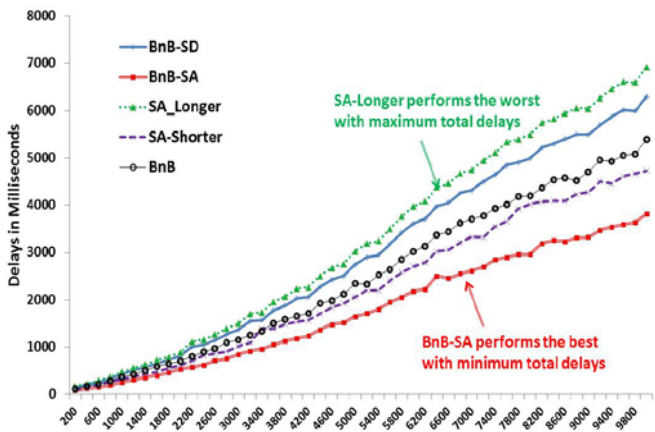

Fig. 6: Total end-to-end delays.

Next, we measure the total resources needed to accommodate all incoming service requests, while there is an upper limit on the total resources, which can be deployed, as explained earlier. BnB-SA again performs the best as seen from Fig. 7, where, the maximum limit is reached when the service requests reach up to 8,400. Since the resource utilization is optimal for BnB-SA, the cost is also minimized for BnB-SA, as shown in Fig. 8.

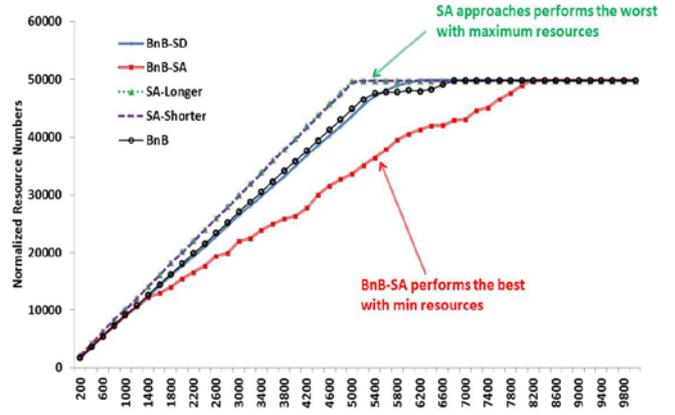

Fig. 7: Total resources required.

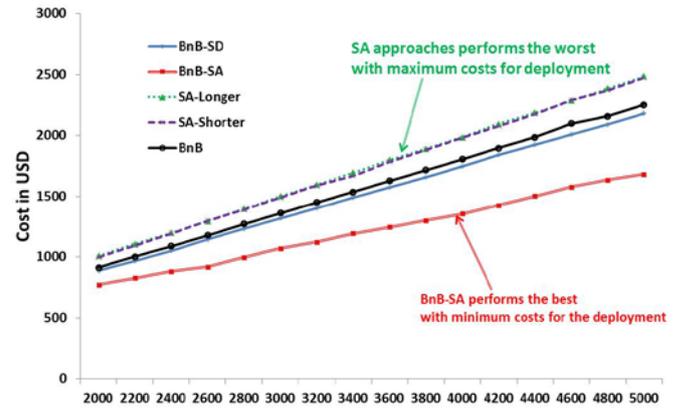

Fig. 8: Total cost of deployment.

TABLE 8. AVERAGE HOPS AS CLOUD NUMBERS VARY.

The superior performance of BnB-SA may be attributed to the fact that in this approach, we are sorting the resource lists in the ascending order of the remaining capacities. It is equivalent to the consolidated approach and as expected, utilizes lowest resources, while satisfying highest service requests. Also, it results in the least number of service instance migrations, as fragmentation of the resources is the least in the consolidated approach. Hence, total end-to-end delays are also smallest with BnB-SA. We also have observed the effect of the total number of clouds deployed in the system as far as overall delays are considered. We argue that it is important to find out the optimal number of clouds for a given topology, since deploying a cloud node incurs significant costs to the cellular service providers. However, a larger number of clouds means proximity of BSs to the clouds, and hence the smaller number of hops and eventually lower delays. On the contrary, it also means more fragmented resources among the clouds (or clouds have smaller capacities) and eventually more service migration instances. Thus, it becomes imperative to find out the optimal number of the clouds to be deployed.

To achieve this, we introduce the delays associated with service migrations, since more fragmented resources mean more relocations. We assume that the average traffic entering per link in the system, with the existing traffic makes the average load per link up to 12 Gbps, which is a general morning peak traffic load [41]. If the average backhaul traffic capacity is assumed to be 20 Gbps, the traffic load comes up to 60%. We find out the average number of hops for a BS to reach the cloud, based on these topologies. Table 7 shows our findings. The complete system has been simulated using Network Simulator 3 (NS3) with the drop-tail queuing system. Numbers of service migration instances are obtained from the heuristic results. Service migration delays are calculated by a simple formula given in [42] as:

*Total Migration Time ≤ Overheads + [(5 × VM Size − 1 ∗ page) /Link Speed]*

TABLE 7. NO. OF HOPS AGAINST NO. OF CLOUDS.

| No. of Clouds | BS/Cloud | Avg. no. of hops |
|---|---|---|
| 1 | 60 | 6 |
| 2 | 30 | 5 |
| 3 | 20 | 4 |
| 4 | 15 | 3 |
| 5 | 12 | 3 |
| 6 | 10 | 2 |
| 7 | 9 | 2 |
| 8 | 8 | 2 |
| 9 | 7 | 1 |
| 10 | 6 | 1 |
| 11 | 5 | 1 |
| 12 | 5 | 1 |
| 13 | 5 | 1 |
| 14 | 4 | 1 |
| 15 | 4 | 1 |

Here, "*page*" is the smallest unit of the data transfer during service migration. Overheads include time to prepare for migrations as well as time for the stop-and-copy stages. VM size is the capacity of the virtual machine on which the service, which is being migrated, is running. For more details about the VM migration details, readers may refer to the works in [42, 43]. Total delays are obtained by adding the link delays as well as the service migration delays. The aim is to find the optimal number of the clouds to be installed during the setup so that the total delays are minimized with the smallest number of clouds deployed. This ensures optimum costs of deployment to the cellular operators as well as minimum total delays.

We plotted the graphs of link delays, service migration delays and total delays in Fig. 9. We used the BnB-SA approach (since it performs the best) for this. Note that as the number of clouds increases, link delays decrease, since the average number of hops decrease as shown in Table 7; however, service migration delays increase, with more fragmentation of the resources. From Fig. 9, we conclude that, for 60 BS and 6000 total service requests, at 60% traffic load, the optimal number of clouds would be 6, since the total delays are minimum at that point.

We also obtained the results using *NS3* for a similar setup. We plotted the graph of the total delays obtained using the BnB-SA approach and *NS3* in Fig. 10 to validate the heuristic results. Note that the simulation results validate the results obtained using heuristic runs since both results overlap with 95% confidence interval. We also plot the graphs for the same at different traffic loads as shown in Fig. 11. Results demonstrate that the optimal number of clouds to be deployed vary as the traffic load varies. For 60 BS and 6000 total service requests, at 60% traffic load, an ideal number of clouds would be 6, however, for the same number of BS and service demands, the number is 12 at 80% traffic load. This may be attributed to the fact that as the traffic load increases, an increase in the number of clouds significantly reduces the link delays. Such delays are significantly higher compared to the service migration delays at higher traffic loads.

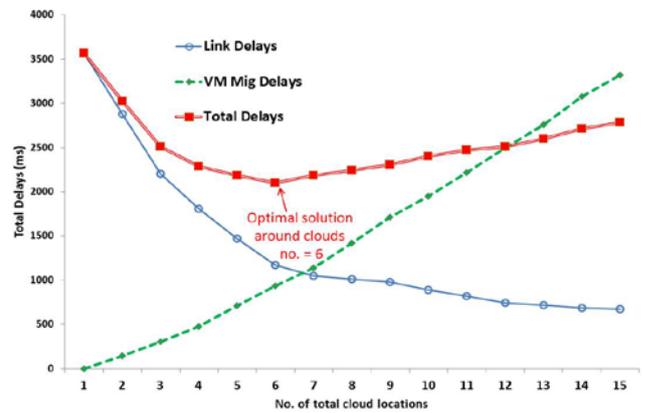

Fig. 9: Optimal number of clouds for minimum total delays.

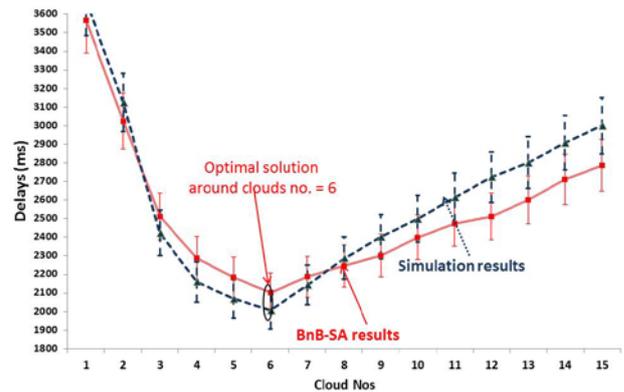

Fig. 10: Variation in delays with cloud numbers.

From the presented results, it is observed that, with the proposed enhancements to BnB approach (BnB-SA), we obtained not only better time complexity, but also improved performance in terms of total delays, total resources utilized and total costs. For the SA, with the smaller number of iterations, time complexity can be reduced significantly. However, the performance is much degraded as compared to the proposed BnB-SA. We also obtained the optimal number of clouds, which needs to be deployed so that the total links delays, as well as the service migration delays, are optimized for the given size of the problem.

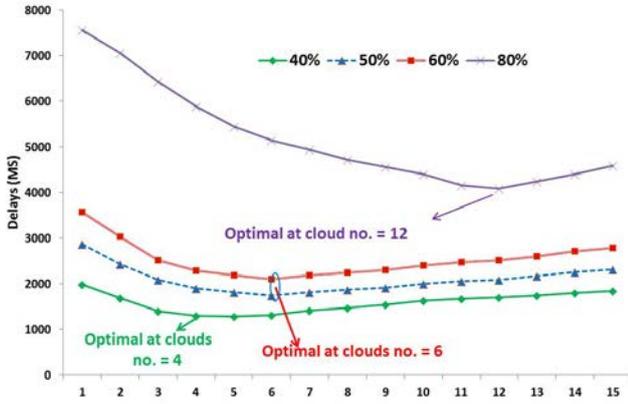

Fig. 11: Optimal cloud numbers at various traffic loads.

The presented solution in this work, however, rely solely on the reactive approach. That is, the placement decisions are made after the service demand request arrives. Making a placement decision and executing that decision may incur significant delays. For example, instantiating some VM as a particular cloud location based on the placement decision may take some time ranging from some seconds to a few minutes. These delays can violate the SLAs. The solution to the problem is a "proactive approach", such as predicting the service demands in advance and have some set of VMs ready at the desired location based on the predictions. Various approaches such as deep learning and machine learning may be used to predict the service demands and proactively preparing placement scheme in advance for the quick response and better utilization of the resources. Also, VMs running in a cloud by using hypervisor technologies available today may incur significant latencies, in the order of micro seconds. Hence, industry and academia are proposing use of micro-services [52] to replace VMs for lesser response time. Future work may include incorporation of micro-services to deploy the service requests in CRANs, instead of virtual machines.

## VI. CONCLUSIONS

In this work, we have addressed the problem of service placement in Cloud Radio Access Networks (CRANs), which is a novel networking paradigm to address the challenges in 5G networks. We argued that to leverage the advantage of this networking paradigm, an efficient service placement scheme is mandatory to meet the stringent latency requirements. We proposed a combinatorial optimization problem to achieve the goal while satisfying other constraints such as cost and capacity constraints.

In addition, we proposed the use of the standard heuristic approaches to solve the larger problems in real time scenarios, which are, branch-and-bound and simulated annealing. We also proposed enhancements to the BnB approach, by sorting the capacity lists in advance. The enhancements reduced the computational complexity resulting in faster deployments of the BBUs and quicker responses to the users. Our goal has been to minimize the total end-to-end delays, while the capacity and cost constraints are met. We demonstrated that one of the proposed enhancements, BnB-SA, performs the best. The computational complexity of the proposed schemes is of the order $O(MlogM)$. However, that of the near-optimal scheme is of the order $O(M^2)$, which is a significant improvement, especially for the CRANs. Using the simulations, we validated the performance improvements by the proposed BnB-SA scheme. We also compared the performance of the proposed schemes against simulated annealing to prove its superiority and observe significant improvements in the solution quality.

With the heuristic implementation, we calculated the optimal number of clouds, which need to be deployed so that the total links delays, as well as the service migration delays, are minimized, while total cloud deployment cost is within the acceptable limits. We also validated our results using the results from NS3 for a similar setup. Future work includes incorporating the proactive approach with the proposed solutions. Various approaches such as deep learning and machine learning may be used to predict the service demands and proactively preparing placement scheme in advance for quick response and better utilization of the resources.

ACKNOWLEDGMENT

This publication was made possible by the NPRP award [NPRP 8-634-1-131] from the Qatar National Research Fund (a member of The Qatar Foundation) and NSF Grant CNS-1718929. The statements made herein are solely the responsibility of the author[s].

TABLE 8
LIST OF ACRONYMS

| Acronym | Description |
|---|---|
| ASP | Application service provider |
| BBU | Baseband unit |
| BnB | Branch and bound |
| BnB-SA | BnB-sorted ascending |
| BnB-SD | BnB-sorted descending |
| BS | Base station |
| CAPEX | Capital expenditures |
| CRAN | Cloud radio access network |
| FFT | Fast Fourier Transform |
| IaaS | Infrastructure as a service |
| ILP | Integer Linear Program |
| IoT | Internet of things |
| ISP | Internet service provider |
| MAC | Media access control |
| MAC-L | MAC lower layer |
| MAC-U | MAC upper layer |
| MIMO | Massive multiple-input multiple-output |
| ms | milliseconds |
| NFV | Network function virtualization |
| NS3 | Network simulator 3 |
| NW | Network |
| OPEX | Operational expenses |
| PHY | Physical |
| QoS | Quality of service |
| RAN | Radio access network |
| RANaas | Radio Access Network-as-a-Service |
| RRH | Remote radio head |
| SA | Simulated Annealing |
| SDN | Software-defined networking |
| SFC | Service Function Chaining |
| SLA | Service level agreement |
| VF | Virtual function |
| VM | Virtual machine |